\documentclass[a4paper,twoside]{article}
\newcommand{\affil}[1]{$^{\rm #1}$}
\usepackage[authoryear]{natbib}
\bibpunct{(}{)}{;}{a}{}{,}
\usepackage{graphicx}
\date{} %Please leave the date blank
\newcommand{\lesssim}{\,\raisebox{-0.4ex}{$\stackrel{<}{\scriptstyle\sim}$}\,}
\newcommand{\gtrsim}{\,\raisebox{-0.4ex}{$\stackrel{>}{\scriptstyle\sim}$}\,}
\title{\large\bf\flushleft Chemical Evolution of the Juvenile Universe}
\author{\parbox{\textwidth}{\flushleft
\vspace{-0.5cm}
{\it G. J. Wasserburg\affil{A,C} and Y.-Z. Qian\affil{B}}\\
\vspace{0.4cm}
{\small \affil{A}\,The Lunatic Asylum, Division of Geological and Planetary Sciences, California Institute of 
        Technology, Pasadena, CA 91125, USA}\\
{\small \affil{B}\,School of Physics and Astronomy, University of Minnesota, Minneapolis, MN 55455, USA}\\
{\small \affil{C}\,Email: gjw@gps.caltech.edu}}}
\begin{document}
\twocolumn[
\begin{changemargin}{.8cm}{.5cm}
\begin{minipage}{.9\textwidth}
\vspace{-1cm}
\maketitle
%
%
%%%%%%%%%%%%%     ABSTRACT    %%%%%%%%%%%%%
%Abstract of no more than 200 words here.
\small{\bf Abstract:} 
Models of average Galactic chemical abundances are in good 
general agreement with observations for ${\rm [Fe/H]}>-1.5$,
but there are gross discrepancies at lower metallicities. 
Only massive stars contribute to the chemical evolution of the ``juvenile universe'' 
corresponding to ${\rm [Fe/H]}\lesssim -1.5$.
If Type II supernovae (SNe II) are the only relevant sources, then the abundances 
in the interstellar medium of the juvenile epoch are simply the sum of different 
SN II contributions. Both low-mass ($\sim 8$--$11\,M_\odot$) and normal 
($\sim 12$--$25\,M_\odot$) SNe II produce neutron stars, which 
have intense neutrino-driven winds in their nascent stages. These winds produce 
elements such as Sr, Y, and Zr through charged-particle reactions (CPR). Such 
elements are often called the ``light $r$-process elements,'' but are 
considered here as products of CPR and not the $r$-process.
The observed absence of production of the low-$A$ elements 
(Na through Zn including Fe) when the true $r$-process elements (Ba and above) 
are produced requires that only low-mass SNe II be the site if the $r$-process 
occurs in SNe II. Normal SNe II produce the CPR elements in addition to the 
low-$A$ elements. This results in a two-component model that is quantitatively 
successful in explaining the abundances of all elements relative to hydrogen
for $-3\lesssim{\rm [Fe/H]}\lesssim-1.5$. 
This model explicitly predicts that ${\rm [Sr/Fe]}\geq-0.32$. 
Recent observations show that there are stars with ${\rm [Sr/Fe]}\lesssim -2$
and ${\rm [Fe/H]}<-3$. This proves that the two-component model is not 
correct and that a third component is necessary to explain the observations. 
The production of CPR elements associated with the formation of neutron stars
requires that the third component must be massive stars ending as black holes. 
It is concluded that stars of $\sim 25$--$50\,M_\odot$ (possibly up to 
$\sim 100\,M_\odot$) are the appropriate candidates. These produce hypernovae 
(HNe) that have very high Fe yields and are observed today. 
Stars of $\sim 140$--$260\,M_\odot$ are completely disrupted upon explosion.
However, they produce an abundance pattern greatly deficient in elements of odd 
atomic numbers, which is not observed, and therefore, they are not considered
as a source here. Using a Salpeter initial mass function, it is shown that HNe are 
a source of Fe that far outweighs normal SNe II, with the former and the latter
contributing $\sim 24\%$ and $\sim 9\%$ of the solar Fe abundance, respectively. 
It follows that the usual assignment of $\sim 1/3$ of the solar Fe abundance to
normal SNe II is not correct. This leads to a simple three-component model 
including low-mass and normal SNe II and HNe, which gives a good description
of essentially all the data for stars with ${\rm [Fe/H]}\lesssim -1.5$. 
We conclude that HNe are more important than normal SNe II in the chemical 
evolution of the low-$A$ elements from Na through Zn (including Fe), in sharp 
distinction to earlier models.

%%%%%%%%%%%%%     KEYWORDS    %%%%%%%%%%%%%
\medskip{\bf Keywords:}
nuclear reactions, nucleosynthesis, abundances --- stars: abundances ---
stars: Population II --- supernovae: general
% Please write all keywords in lower case. PASA uses the
% standard list of subject headings adopted by The Astrophysical Journal
% and available from http://www.journals.uchicago.edu/ApJ/keywords_text.html.
% Keywords are separated by em-dashes, i.e. ---

%%%%%%%%DO NOT EDIT%%%%%%%%%%%%
\medskip
\medskip
\end{minipage}
\end{changemargin}
]
\small
%%%%%%%%EDIT FROM HERE%%%%%%%%%%%%

\section{Introduction}
\begin{figure}[h]
\begin{center}
\includegraphics[scale=0.38]{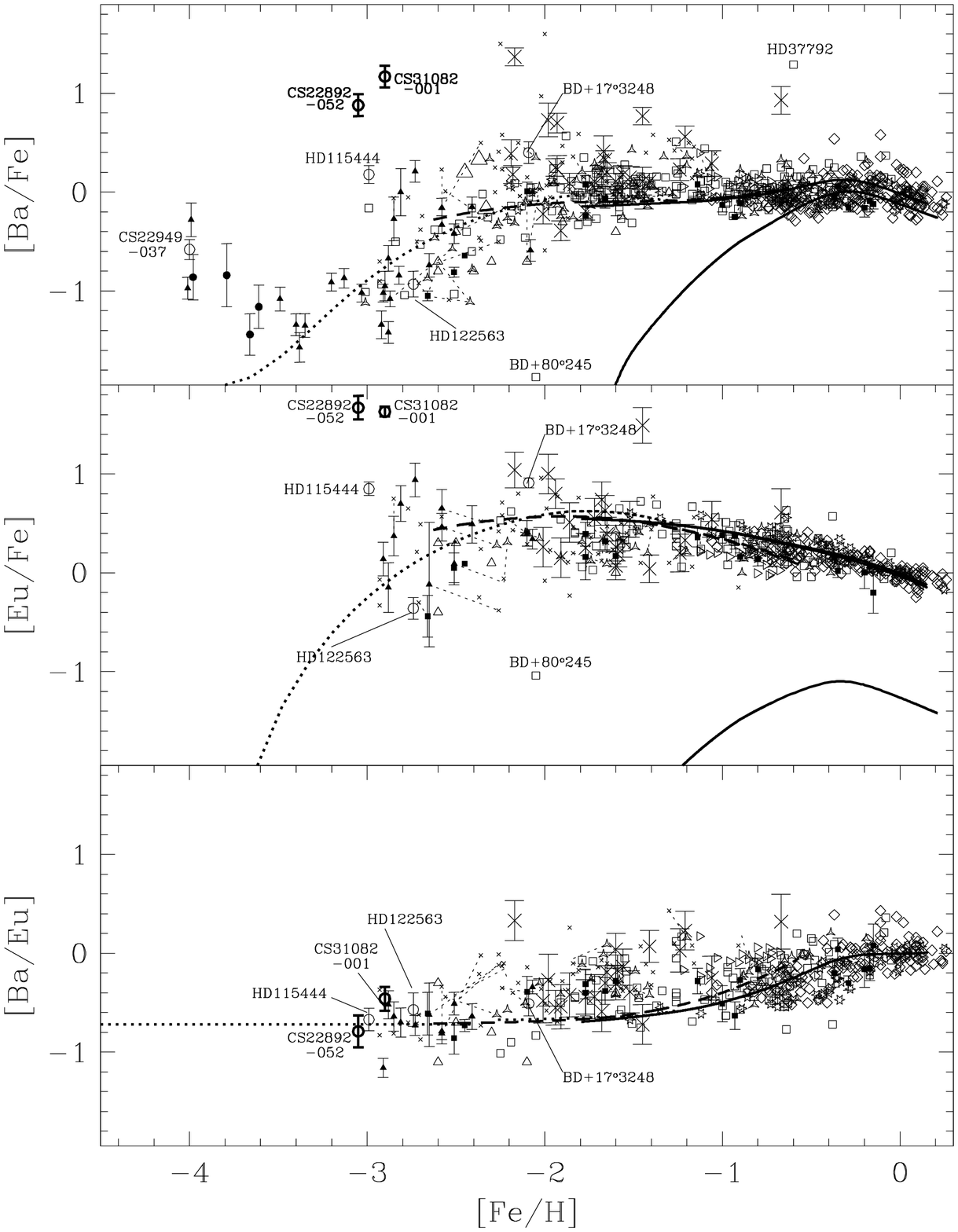}
\caption{Comparison of the results from a Galactic chemical evolution model with
the data on [Ba/Fe], [Eu/Fe], and [Ba/Eu] vs. [Fe/H] (taken from Figure~5 in 
\citealt{trav04}; see that reference for the data sources. Reproduced by permission
of the AAS). The solid, dashed, and dotted curves
show the evolution of [Ba/Fe], [Eu/Fe], and [Ba/Eu] with [Fe/H] for stars in the thin disk, 
thick disk, and halo, respectively. The $s$-process contributions to Ba and Eu are shown 
by the thick solid curves in the top two panels. 
The regime of ${\rm [Fe/H]}\protect\lesssim -1.5$ is
taken to represent the ``juvenile'' universe.\label{fig1}}
\end{center}
\end{figure}

%Please see the PASA Style Guide for help with correct layout for your manuscript.
%Examples of tables and figures are given below.
The problem of the ``chemical evolution of the Galaxy'' has attracted many workers over the
last several decades. The models that seek to address this evolution are, for the most part,
focused on contributions from Type II supernovae (SNe II), SNe Ia, and asymptotic
giant branch (AGB) stars. The problem is complex as it depends on
the elemental yields of all the diverse sources. For example, both the ``weak'' and ``main'' 
$s$-process aspects as well as the $r$-process contributions must be included in such 
calculations in order to treat the neutron-capture elements (e.g., \citealt{trav04}). 
The complication also comes
from the large uncertainties in the frequencies of occurrence for the sources and the problem
of mixing, which may be local or between different regions of the Galaxy (e.g., disk and halo).
The results from these chemical evolution studies, while very model dependent, give a good
description of the general average abundance patterns as a function of 
[Fe/H]~$=\log({\rm Fe/H})-\log({\rm Fe/H})_\odot$ for disk stars. 
The results from one such study are shown in Figure~\ref{fig1} (taken from Figure~5 in 
\citealt{trav04}). As can be seen, the solid and dashed curves give 
a good description of the evolution of [Ba/Fe], [Eu/Fe], and [Ba/Eu] for stars with 
[Fe/H]~$>-1.5$ in the thin and thick disk, respectively.
The thick solid curves in the top two panels showing only the $s$-process contributions
to Ba and Eu indicate that the onset of major $s$-process contributions is at [Fe/H]~$\sim -1$ 
and that the $s$-process contributes very little Eu.

The analyses illustrated in Figure~\ref{fig1} are for ``average abundances'' and do not
predict the abundance (E/H) of element E relative to hydrogen in an individual star.
Inspection of Figure~\ref{fig1} shows that this description of abundances
fails for [Fe/H]~$\lesssim -1.5$, the regime of which represents the ``halo'' phase
in the interconnected evolution of the disk and the halo
(see the dotted curves in Figure~\ref{fig1}).
It is this regime that is the focus of our interest. We consider that 
[Fe/H]~$\lesssim -1.5$ corresponds to
a ``juvenile'' universe. It will be shown that data in this juvenile
regime with far fewer contributing stellar sources lead to a clearer understanding of the 
nature and types of these sources in spite of the very large discrepancies between the
model of average abundances and the data shown in Figure~\ref{fig1}.

In all of the Galactic chemical evolution (GCE) studies, success in reproducing the solar 
abundances is taken as a measure of validity of the approach. When 
an element receives contributions from multiple sources, to correctly account for its solar 
abundance requires the identification of all the important sources and the calculation of the 
relative contributions of these sources. In this case, failure to include all the relevant sources
results in erroneous attribution to the sources that are included 
in the GCE model to reproduce the solar abundances. When the contributing stellar sources 
evolve on very different timescales, observations covering a wide range of [Fe/H] can help
in estimating the relative importance of these sources. For example, Fe production in
SNe II associated with rapidly-evolving massive stars and in SNe Ia associated with 
slowly-evolving low-mass stars is well established by both observation and theory. 
In contrast, O is produced in SNe II but not in SNe Ia. As a result, there is a general trend
for [O/Fe] to decrease as [Fe/H] increases. Using the data on [O/Fe] vs. [Fe/H], 
\cite{timmes} estimated that $\sim 1/3$ to 1/2 of the solar Fe abundance is produced by
SNe II. However, it will be shown that this attribution to SNe II is far too large and that another
massive stellar source for Fe and other elements of ``low'' mass numbers
(low-$A$ elements from Na through Zn including Fe with mass numbers $A\sim 23$--70) is 
in play. It will be shown that 
the effects of this additional source cannot be discerned based on timescales for 
stellar evolution alone, but are exhibited by the data on three groups of elements 
represented by Fe, Sr, and Ba, respectively, for metal-poor stars with [Fe/H]~$\lesssim-1.5$
formed in the juvenile universe. 

\begin{figure}[h]
\begin{center}
\includegraphics[scale=0.38]{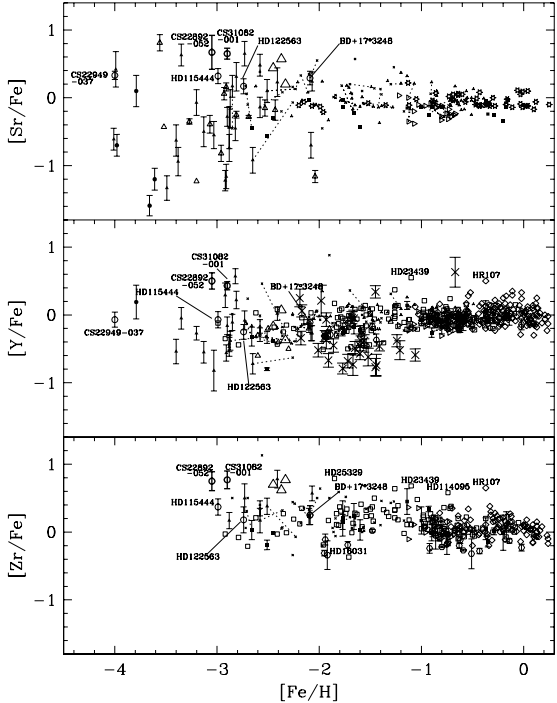}
\caption{Data on [Sr/Fe], [Y/Fe], and [Zr/Fe] vs. [Fe/H] (taken from Figure~4 in 
\citealt{trav04}; see that reference for the data sources. Reproduced by permission
of the AAS). Note the wide scatter of data for [Fe/H]~$\protect\lesssim -1.5$, especially
for [Fe/H]~$\protect\lesssim -3$.\label{fig2}}
\end{center}
\end{figure}

The increase of [Fe/H] is usually taken as a measure of the passage of time. However,
very low [Fe/H] values cannot give precise timing for the chemical enrichment but only 
indicate an early epoch during which few enrichment events occurred in a local region.
It is widely recognized that metal-poor stars, especially those with [Fe/H]~$\lesssim -3$,  
sample grossly inhomogeneous mixtures of the products from various massive stellar 
sources. This inhomogeneity is clearly demonstrated by the large scatter of $\gtrsim 2$~dex
in [Ba/Fe] and [Eu/Fe] at [Fe/H]~$\sim -3$ in contrast to the reasonably well-defined trends 
at [Fe/H]~$>-1.5$ shown in Figure~\ref{fig1}. The same conclusion is also reached from
the data on [Sr/Fe], [Y/Fe], and [Zr/Fe] vs. [Fe/H] shown in Figure~\ref{fig2} (taken from
Figure~4 in \citealt{trav04}). In particular, there is wide scatter in [Sr/Fe] at any specific
[Fe/H] for $-4\lesssim{\rm [Fe/H]}\lesssim -3$. Clearly, to account for the scatter shown
in Figures~\ref{fig1} and \ref{fig2} requires multiple massive stellar sources with 
greatly-varying yields of Sr, Y, Zr, Ba, and Eu relative to Fe. It will be shown that
some elements such as Ba and Eu are never co-produced with Fe and others such as
Sr, Y, and Zr are produced sometimes with and sometimes without Fe. 

Professor Roberto Gallino has persistently emphasized that understanding the production 
of Sr, Y and Zr is a key to understanding the evolution of the elements. Their possible origin 
in the $r$-process and $s$-process is problematic. We have followed Roberto's guidance in 
this important problem, 
but not in a direction that he would approve, and report our findings here. 
The purpose is to make certain that
his blood pressure is kept sufficiently elevated to support his ``vital signs'' over the next two 
decades. Rather than using estimates of elemental yields
from stellar models, we focus on the observed abundances at low metallicities 
where contributions to the interstellar medium (ISM) from SNe Ia and AGB stars should be 
small and the dominant contributions come from massive stars with short lifetimes. Our approach is to
use the elemental abundances observed in metal-poor stars as a guide to the stellar sources of nuclei,
in particular Sr, Y, and Zr, as well as Fe, Ba, and Eu. The abundances of these elements
observed in selected stars are used as the templates of nucleosynthesis for the identified sources.
Before we present our latest results, we give
a brief review of some earlier works in connection with or parallel to this approach.

\section{Review of earlier works}
Since the seminal works of \cite{b2fh} and \cite{cameron}, elements heavier than the Fe group 
have been considered as produced predominantly by the $r$-process and $s$-process.
The solar abundances, which are largely based on measurements of isotopic and chemical 
abundances in meteorites, are the observational basis for de-convoluting the $r$-process and
$s$-process components in the ``bulk solar'' abundance data. The net solar abundance of an 
element E is 
\begin{equation}
N_\odot({\rm E})= N_{\odot,r}({\rm E})+N_{\odot,s}({\rm E}),
\end{equation}
where $N_{\odot,r}({\rm E})$ and $N_{\odot,r}({\rm E})$ are the contributions from the
$r$-process and $s$-process, respectively. The $s$-process contributions 
($s$-contributions) have been
studied by many workers. This allows the $r$-contributions to be inferred by
subtracting the $s$-contributions from the net solar abundances (e.g.,
\citealt{kappeler,arlandini}). In cases where the $s$-contributions
are large or dominate, there is considerable uncertainty in inferring the $r$-contributions.
This is true for Sr,Y, Zr, and Ba. Recent advances in the study of the $s$-process
have been largely due to the exquisite experimental work on neutron-capture cross sections 
at stellar energies by F. K\"appeler and his colleagues at Karlsruhe and to the thorough and 
deep analyses of the $s$-process in AGB stars (with assumptions about the ``{$^{13}$C}
pocket'' as the neutron source) by R. Gallino and his colleagues at Torino 
[see e.g., \cite{busso} for a review].

The matter of $r$-contributions is particularly ill-defined as the ``$r$-process'' site is assigned 
to SNe II but models have not succeeded in finding a suitable stellar environment with adequate 
neutron flux. It was anticipated that the right conditions for the $r$-process would be found in
the neutrino-driven winds from a nascent neutron star (e.g., \citealt{woosley94}), but at present, 
this approach is without success [see e.g., \cite{qian} for a review]. Parametrized models 
assuming an adequate neutron source are capable of fitting the inferred solar $r$-process
abundance pattern [solar $r$-pattern, see e.g., \cite{kratz,meyer,freiburghaus}]. These approaches 
calculate, in some detail, the relative $r$-process yields and loosely associate them with
SNe II resulting from core collapse. However, such models are without a direct consequential 
relationship to stellar evolution and explosion. In particular, there is no basis in such calculations 
for the decoupling of the ``heavy'' $r$-process elements ($r$-elements) such as Ba and Eu
from the low-$A$ elements such as Fe (see below). Similar parametrized calculations 
were also performed to fit the abundances of the $r$-elements in metal-poor stars 
(e.g., \citealt{montes,farouqi}).

New insights into the $r$-process were gotten from investigation of short-lived nuclei in the 
early solar system. Both $^{129}$I and $^{182}$Hf are produced predominantly by the 
$r$-process and their lifetimes are similar. The discrepancy found between their abundances 
in meteorites led to the proposal by \cite{wbg} that there was not ``an $r$-process'' but 
that there had to be two (or more) $r$-processes to explain the meteoritic data. 
These workers proposed that the sites for producing the ``heavy'' and ``light'' $r$-elements
with $A>130$ and $A\lesssim 130$, respectively, were different, with the source ($H$) for
heavy $r$-elements occurring at a high frequency and that ($L$) for the light ones at
a low frequency. Further, they predicted that, at low metallicities where fewer sources might 
contribute to the ISM from which stars formed, the relative abundances of heavy and light 
$r$-elements should show a scatter. This model of diverse $r$-process sources was 
extensively developed by \cite{qw00} with the $H$ source predicted to have
a very low yield ratio of light $r$-elements relative to the heavy ones. 
It was found by \cite{sneden} that 
the abundance of the light $r$-element Ag relative to the heavy $r$-element Eu 
in the ultra-metal-poor star CS~22892--052 was significantly lower compared with the
solar $r$-pattern, but not nearly as low as initially predicted for the $H$ source by 
\cite{qw00} (see also \citealt{qvw}).

The timescales for stellar evolution have been 
well established. It is clear that during the first 
$\sim 10^9$~yr after the big bang, the stellar
sources contributing to the ISM 
(and the intergalactic medium, IGM) 
must have masses of $M\gtrsim 3\,M_\odot$,
and the dominant sources must be massive stars 
of $M\gtrsim 8\,M_\odot$ with rapid evolutionary 
timescales of $\ll 10^9$~yr. We define this domain 
as  the  Òjuvenile universeÓ and consider SNe II
and other massive stars as the major sources. 
Note that SNe Ia are associated with
the evolution of low-mass stars in binaries and 
thus cannot have been major early polluters. 
Taking $\sim 2/3$ of the solar Fe abundance to be 
contributed by SNe Ia and the rest by
all massive stellar sources over a period of 
$\sim 10^{10}$~yr prior to the
formation of the solar system, we estimate that
a metallicity of [Fe/H]~$\sim\log(1/30)\sim-1.5$
is reached by the end of the first $\sim 10^9$~yr.
Thus the juvenile universe corresponds to
[Fe/H]~$\lesssim -1.5$ (where the $s$-contributions
from AGB stars to the ISM are also negligible; see Figure~\ref{fig1}).
This is precisely the region 
where the broad-brush GCE models have failed 
(see Figure~\ref{fig1}). 

\begin{figure}
\begin{center}
\includegraphics[scale=0.42]{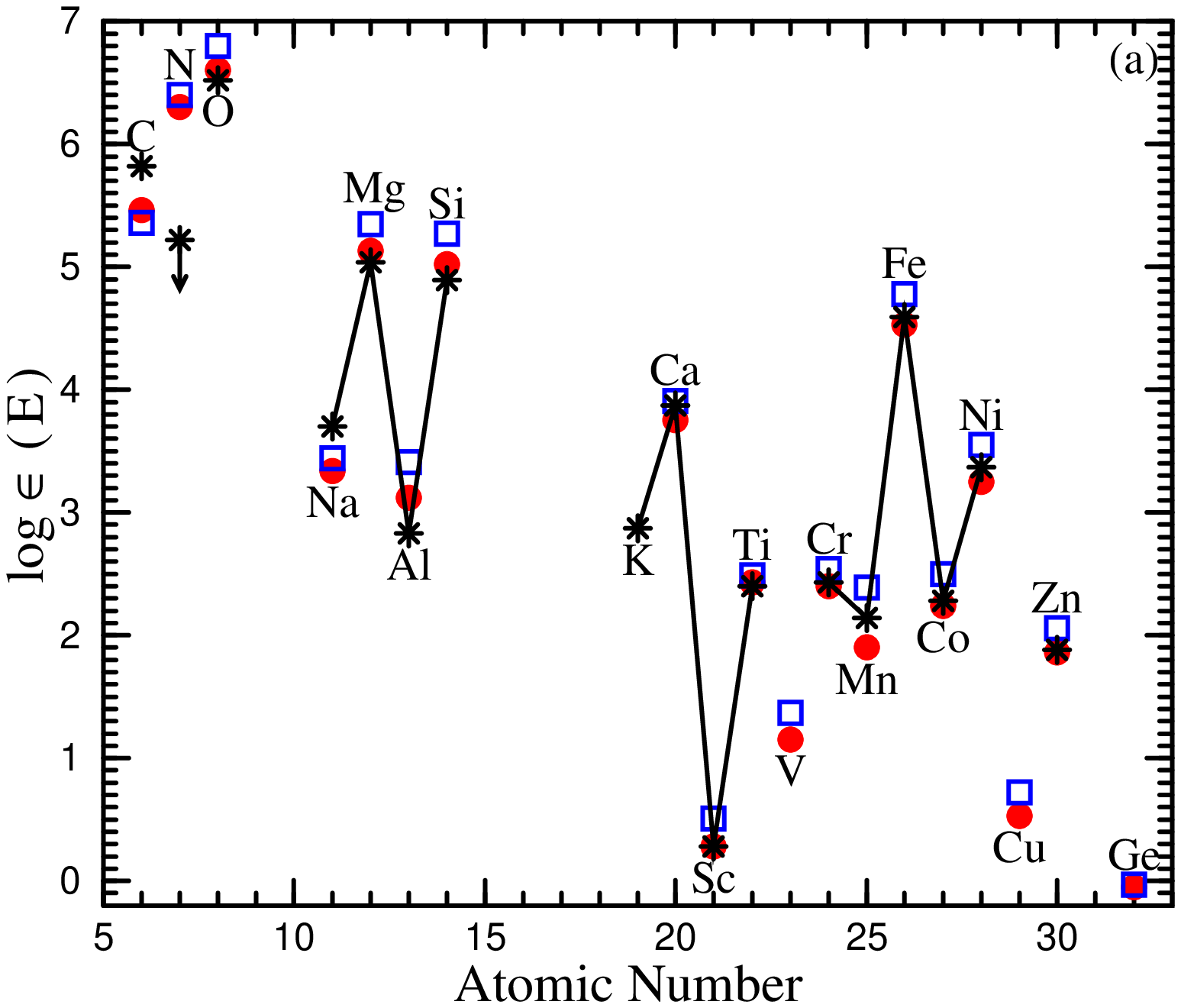}
\includegraphics[scale=0.42]{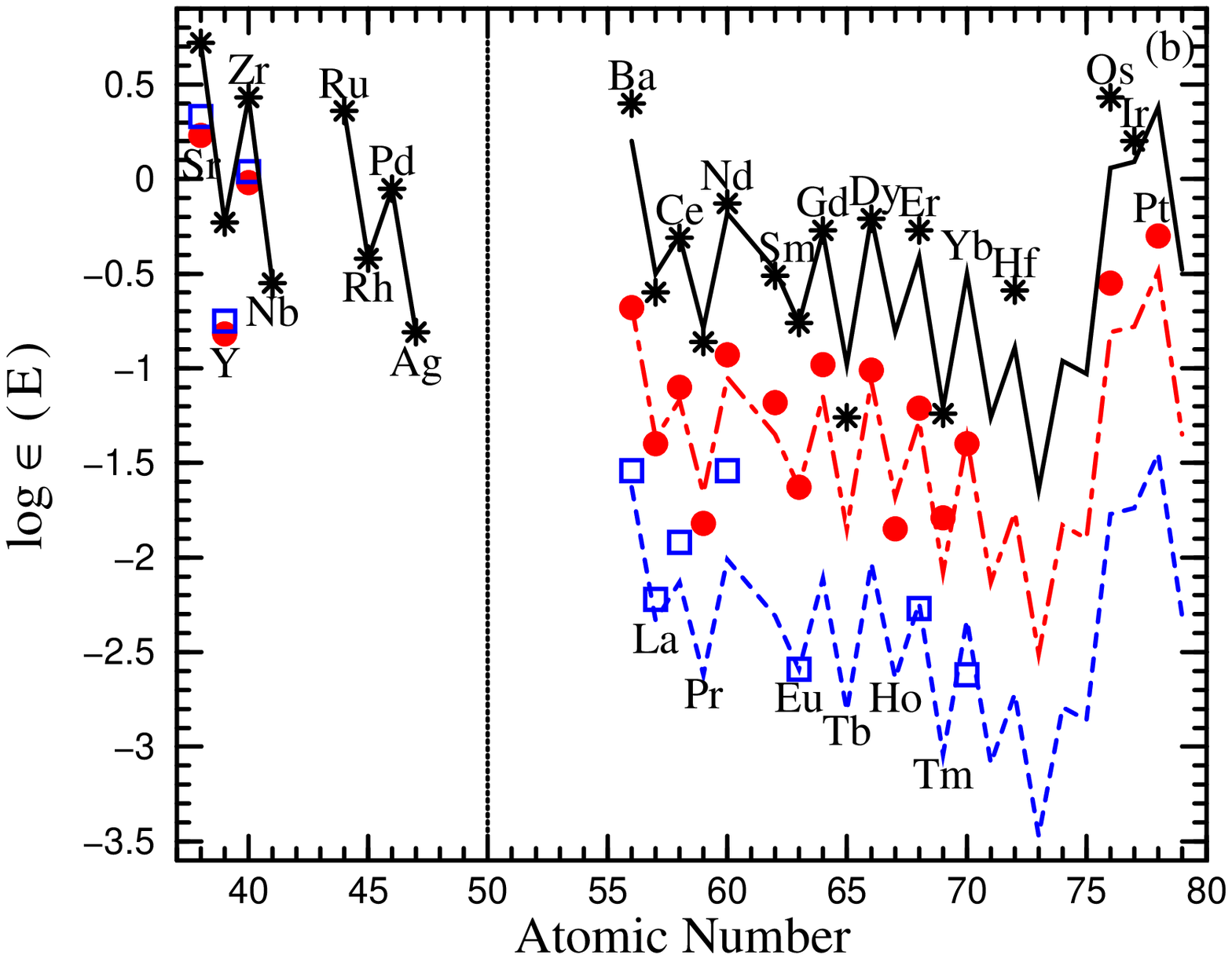}
\caption{Data on CS~31082--001 (asterisks;
\citealt{hill}), HD~115444 (filled circles), and HD~122563 (squares; \citealt{westin})
with [Fe/H]~$=-2.9$, $-2.99$, and $-2.74$, respectively.
(a) The values of $\log\epsilon({\rm E})\equiv\log({\rm E/H})+12$
for the elements from C through Ge. The data
on CS~31082--001 are connected by solid line segments as a guide.
Note that the available abundances for the low-$A$ elements from Na through Zn 
are almost indistinguishable
for the three stars. (b) The $\log\epsilon$ values for 
the elements from Sr through Pt. The data for
CS~31082--001 in the region to the left of the 
vertical line are again connected by solid line segments as a guide. 
In the region to the right of the vertical line, the data on
the heavy $r$-elements are compared with the solid, dot-dashed, and 
dashed curves, which are the solar $r$-pattern \citep{arlandini} translated to
pass through the Eu data for CS~31082--001, HD~115444, and
HD~122563, respectively. Note the general agreement between the
data and the solid and dot-dashed curves. There is a range of 
$\sim 2$~dex in the abundances of the heavy $r$-elements for the
three stars shown. Combined with their nearly identical abundances
of the low-$A$ elements, this shows that the production of the heavy
$r$-elements is independent of the production of the low-$A$ elements.\label{fig3}}
\end{center}
\end{figure}

The main facts that are actually known about the
juvenile universe are the observed abundances in 
metal-poor stars residing in the Galactic halo. 
The observational studies of these stars have blossomed to 
produce a considerable database of high-quality elemental 
abundances, which has been the guide to all advances 
in the field. In particular, the observational data show that
the production of the heavy $r$-elements (Ba and above)
are independent of the production of the low-$A$ elements
from Na through Zn including Fe (see Figure~\ref{fig3}).
This means that associating the $r$-process with SNe II
requires that such SNe II cannot produce much Fe. This
recognition led us to conclude that if the $r$-process occurs
in SNe II,  the site must be low-mass ($\sim 8$--$11\,M_\odot$)
SNe II \citep{qw02,qw03}. Further, the data showed that
the abundances of all heavy $r$-elements closely follow
the solar $r$-pattern in general. Thus this pattern appears
to be relatively robust, although there are deviations in some
cases. The above two observational facts are the basis of
all our discussions.

Using the available data and considering
that the diversity of stellar sources must be quite 
restricted, efforts were made to identify the yield 
templates of potential significant sources (e.g., 
\citealt{qw01,qw02}).  
The paper titled ``A Model for Abundances in Metal-Poor Stars''
\citep{qw01} argued that the elemental abundances 
relative to hydrogen in metal-poor stars can be explained  
by two kinds of SNe II ($H$ and $L$) and the ``Population III''
(Pop III) very massive stars 
(VMS), which were supposed to occur only in the earliest epochs 
and provided a ``prompt inventory'' ($P$-inventory) of metals. 
The yield templates of all three sources were obtained from 
the solar $r$-contributions and the abundances in two selected 
stars with [Fe/H]~$\approx -3$
but with very different abundances of heavy $r$-elements.
It was found that
a good match to the abundances of Sr, Y, Zr, and Ba (relative
to hydrogen) could be obtained for many stars 
if the standard solar $r$-contributions 
of these elements were substantially increased.
This suggested that it was necessary to revise the solar $r$-contributions, 
and hence the solar $s$-contributions, of Sr, Y, Zr, and Ba reported by 
\cite{kappeler} and \cite{arlandini}.

A study on GCE of Sr, Y, and Zr by \cite{trav04} found that
if data on metal-poor stars were included in the evolution calculation, 
then this required increases in the solar $r$-contributions of Sr, Y, and Zr 
in quantitative accord with the results of \cite{qw01}. 
These revised solar $r$-contributions are essentially what we use at present. 
\cite{trav04} further proposed that some lighter element primary process (LEPP),
possibly a variant of the $r$-process, had to exist as an additional source
for Sr, Y, and Zr.

The general approach taken by us is that stars are reliable adding machines 
operated by the diverse sources contributing to the ISM from which they formed.
Certain ``rules'' for the stellar adding machines found by 
\cite{qw01} turned out to be misleading. There was an apparent sharp increase
in abundances of heavy $r$-elements at [Fe/H]~$\sim-3$. It was assumed that
this metallicity represented a ``baseline'' enrichment (the $P$-inventory) due to 
production by VMS \citep{qw02}. It was shown later that this baseline
enrichment could be reached rapidly through production by SNe II inside halos
of sufficient mass that could gravitationally bind the SN II debris \citep{qw04}.
Thus there is no need for the $P$-inventory proposed earlier.

As the field developed, we found that essentially all of the observed 
abundances for a large number of elements (relative to hydrogen) in stars
with $-3\lesssim{\rm [Fe/H]}\lesssim -1.5$ could be explained by a mixture of 
two components, $H$ and $L$, with yield patterns taken from two
template stars \citep{qw07}. The $H$-contributions are measured by 
the abundance of
a heavy $r$-element such as Eu (or Ba) and the $L$-contributions by
that of Fe. The study of $r$-process models in conjunction with the 
observational data on metal-poor stars also led \cite{montes} to the 
conclusion that a two-component model can explain the available data. 
These workers explored the possibility that a variant of the $r$-process 
could be the LEPP source for Sr, Y,  and Zr proposed by \cite{trav04}. 

The results of the two-component model led to some surprising conclusions: 
(1) the abundances of a large number of elements relative to 
hydrogen can be calculated with considerable reliability for essentially all stars 
with $-3\lesssim{\rm [Fe/H]}\lesssim -1.5$ using the assumed $H$ and $L$ 
yield templates, (2) the production of heavy $r$-elements 
was decoupled from that of the low-$A$ elements from Na through Zn
including Fe, and (3) the abundance patterns of the
low-$A$ elements were essentially the same for all stars  
with $-3\lesssim{\rm [Fe/H]}\lesssim -1.5$ with only few exceptions. 
These rules led to the conclusion that normal SNe II of $\sim 12$--$25\,M_\odot$, 
which produce Fe, cannot be the source for the heavy $r$-elements. 
This then restricted the possible source for these elements to
low-mass SNe II of $\sim 8$--$11\,M_\odot$ with very little Fe production 
[see e.g., \cite{ning} for such an $r$-process model]. 
Rule (3) was particularly surprising as the yields of SNe II of different masses 
are known to be quite variable (e.g., \citealt{whw}). 
Hence, with few stellar sources contributing 
to a local ISM at very low [Fe/H], there should be substantial scatter in the 
abundance patterns of the low-$A$ elements in metal-poor stars. 
This is not the case in general.

The matter was brought to sharp focus by the following: (1) the assignment of the 
heavy $r$-elements to low-mass SNe II,
(2) both low-mass and normal SNe II produce neutron
stars as shown by theoretical models (e.g., \citealt{nomoto,whw}), (3)
many of the so-called light $r$-elements such
as Sr, Y, and Zr are readily produced in the neutrino-driven winds from 
nascent neutron stars (e.g., \citealt{wh92}) and not in a true $r$-process,
and (4) there was no sound basis for relating the high 
neutron fluxes required for production of the heavy $r$-elements to 
a neutrino-driven wind directly.
This then provided qualitative justification for taking the relative
yields of the light to heavy $r$-elements for the $H$ source 
(low-mass SNe II) and those
of the light $r$-elements to Fe for the $L$ source (normal SNe II)
from the abundances in two template stars. The so-called light $r$-elements
such as Sr, Y, and Zr are now specifically attributed to charged-particle
reactions (CPR) in the neutrino-driven winds and thus are not considered
to be true $r$-process elements. The sources proposed for the
two-component model are summarized in Table~\ref{tab1}.

\begin{table}[h]
\begin{center}
\caption{Two-Component Model}\label{tab1}
\begin{tabular}{lccc}
\hline & low-$A$& CPR& heavy \\
&elements&elements&$r$-elements\\
\hline 
low-mass&no&yes&yes\\
SNe II ($H$)&&&\\
normal&yes&yes&no\\
SNe II ($L$)&&&\\
\hline
\end{tabular}
\end{center}
\end{table}

The two-component model had clear predictions about the possible 
values of [Sr/Fe], [Y/Fe], and [Zr/Fe] that could be observed \citep{qw07}.
In particular, as normal SNe II ($L$) produce both Sr and Fe
while low-mass SNe II ($H$) produce Sr but no Fe
(see Table~\ref{tab1}), the lowest value of Sr/Fe predicted
by this model is fixed by the yield ratio (Sr/Fe)$_L$ for normal SNe II.
This lower limit is [Sr/Fe]~$\geq -0.32$.
Using the $H$ and $L$ yield templates, the abundances of all elements
in any star with [Fe/H]~$\lesssim -1.5$ can be calculated from the
observed abundances of an heavy $r$-element such as Eu (or Ba)
and a low-$A$ element such as Fe. This model led to excellent
predictions for many stars with $-3\lesssim{\rm [Fe/H]}\lesssim-1.5$.
However, more extensive data at [Fe/H]~$<-3$ showed that there
was a clear violation of the lower limit on [Sr/Fe].
It is the failure of this prediction by the two-component model that has led to 
a clearer view of the contributors to the chemical evolution of both the juvenile 
universe and the subsequent epoch. It will be shown that massive stars of
$\sim 25$--$50\,M_\odot$ (possibly up to $\sim 100\,M_\odot$)
are very important players throughout the
history of the universe. This is the focus of the present work.

\section{Three-component model}
The high-resolution (e.g., \citealt{johnson02,honda04,aoki05,francois07,cohen08}) 
and medium-resolution \citep{barklem05} observations of elemental abundances 
in a large number of low-metallicity stars in the Galactic halo 
[and a single star in a dwarf galaxy \citep{fulbright04}] now provide 
a data base for determining the nature of the stellar sources contributing 
to the ISM/IGM at metallicities of 
$-5.5<{\rm [Fe/H]}\lesssim-1.5$. These data taken in conjunction with stellar models appear 
to define the massive stars active in the juvenile universe. This changes our views of 
what may be Pop III stars and what stellar types are continuing 
contributors through the present epoch. 

\begin{figure}
\begin{center}
\includegraphics[angle=270,scale=0.3]{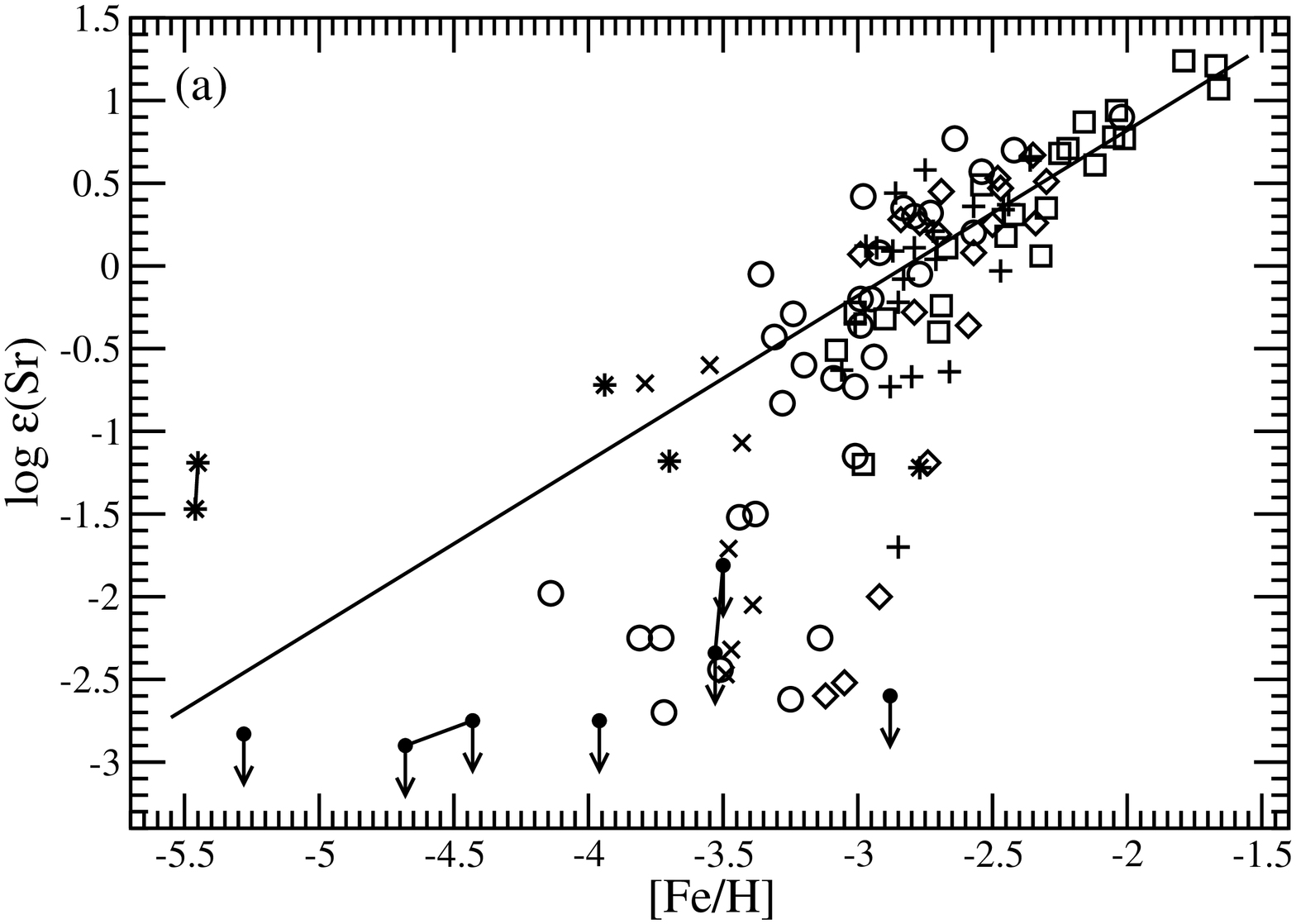}
\includegraphics[angle=270,scale=0.3]{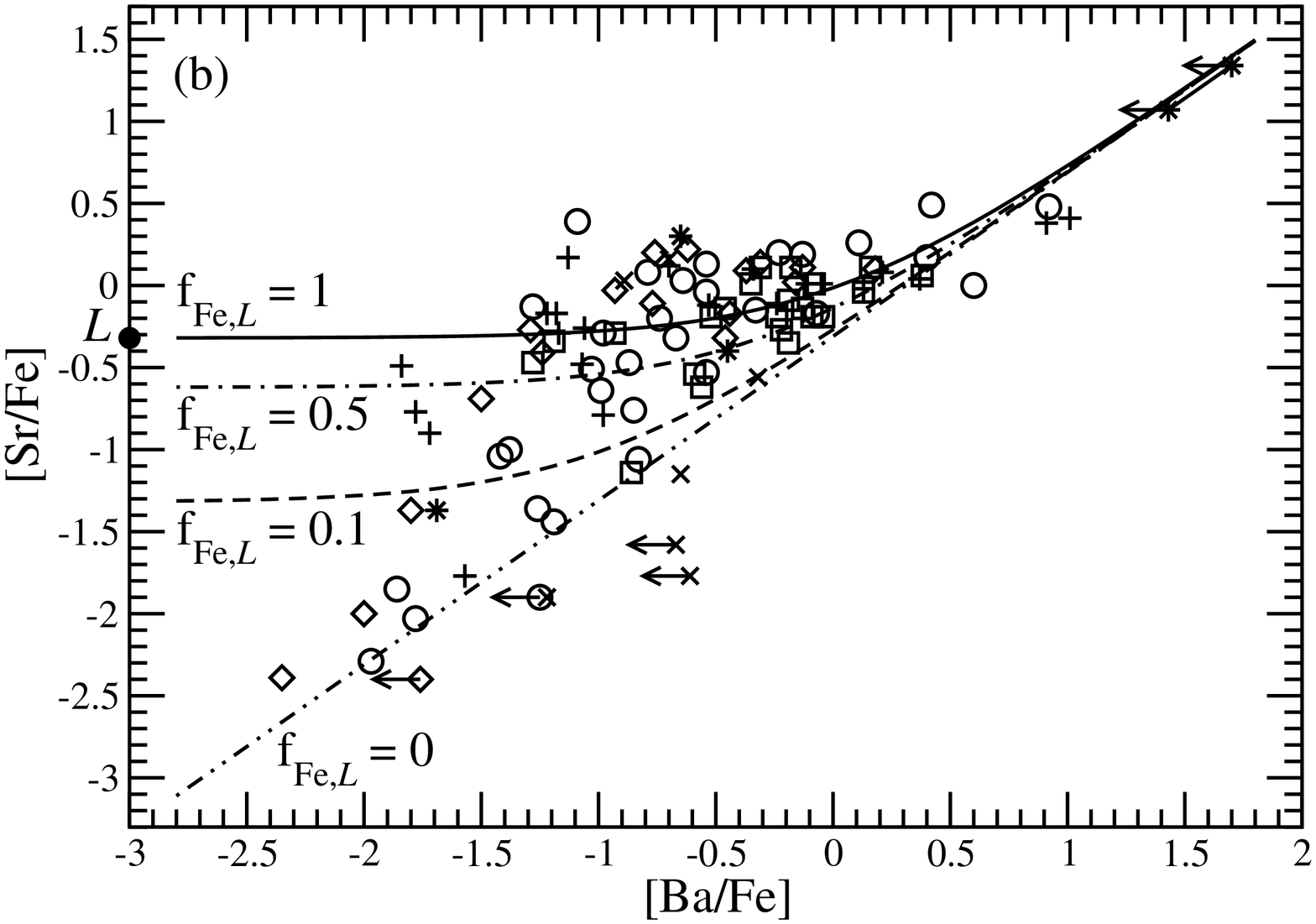}
\caption{(a) High-resolution data on $\log\epsilon({\rm Sr})$ vs. [Fe/H]. The solid line is 
calculated from the two-component model for a well-mixed ISM/IGM. Note that there is 
a great deficiency of Sr for many sample stars with ${\rm [Fe/H]}\protect\lesssim-3$. 
It is evident that a source producing Fe and no Sr is required. 
(b) Evolution of [Sr/Fe] with [Ba/Fe] for
the data shown in (a). The curves correspond to different fractions $f_{{\rm Fe},L}$ of 
Fe due to the $L$ source (normal SNe II). 
Note the data points lying on the $f_{{\rm Fe},L}=0$ curve and those above
the $f_{{\rm Fe},L}=1$ curve. Details for these and the subsequent figures 
as well as the data sources
can be found in \cite{qw08}.\label{fig4}}
\end{center}
\end{figure}

As mentioned in Section~2, the two-component model predicts [Sr/Fe]~$\geq -0.32$
for all stars with [Fe/H]~$\lesssim -1.5$ \citep{qw07}.
This rule appears to be well followed for ${\rm [Fe/H]}>-3$. However, a serious problem
with this model arises below ${\rm [Fe/H]}\sim -3$. In this domain the extended database
shows that there is a gross deficiency of Sr (and other CPR elements) relative to Fe
(see Figure~\ref{fig4}a). It follows that a third component in addition to 
low-mass and normal SNe II is required to account for all the abundance data.
Further, if Sr, Y, and Zr are CPR nuclei, then this third component 
must be a massive stellar source of Fe leaving behind a black hole instead of 
a neutron star that can produce the CPR nuclei in the neutrino-driven winds. 

The effects of all three stellar sources are most clearly seen if we consider the 
relationship of Sr (a CPR element in our model) in conjunction with Ba 
(a true $r$-element at low metallicities) and Fe. Using the yield ratios (Sr/Ba)$_H$ of the
$H$ source (low-mass SNe II) and (Sr/Fe)$_L$ of the $L$ source (normal SNe II)
we obtain: 
\begin{equation}
\left(\frac{\rm Sr}{\rm H}\right)=\left(\frac{\rm Sr}{\rm Ba}\right)_H\left(\frac{\rm Ba}{\rm H}\right)
+\left(\frac{\rm Sr}{\rm Fe}\right)_L\left(\frac{\rm Fe}{\rm H}\right)f_{{\rm Fe},L},  
\end{equation}
where $f_{{\rm Fe},L}$ is the fraction of Fe from the $L$ source 
(the two-component model corresponds to $f_{{\rm Fe},L}=1$). The above 
equation can be rewritten as:
\begin{equation}
{\rm [Sr/Fe]}=\log\left(10^{{\rm [Sr/Ba]}_H+{\rm [Ba/Fe]}}+ f_{{\rm Fe},L}\times 
10^{{\rm [Sr/Fe]}_L}\right). 
\label{eq3}
\end{equation}
 
The evolution of [Sr/Fe] with [Ba/Fe] for the high-resolution data shown in 
Figure~\ref{fig4}a is exhibited in Figure~\ref{fig4}b along with the curves 
representing Equation~(\ref{eq3}) for $f_{{\rm Fe},L}=0$, 0.1, 0.5, and 1
(using [Sr/Ba]$_H=-0.31$ and [Sr/Fe]$_L=-0.32$).
Similar results are found for the medium-resolution data. 
While there appears to be a clustering of data in the neighborhood of 
$f_{{\rm Fe},L}=1$ corresponding to Fe contributions exclusively from the $L$ 
source in Figure~\ref{fig4}b, there is a substantial fraction of the data lying 
down to $f_{{\rm Fe},L}=0$. This requires an Fe source not related to normal
SNe II and clearly shows that the preponderance of the Fe in many sample 
stars is from this third source that produces no Sr. 
Essentially the same results shown for [Sr/Fe]
vs. [Ba/Fe] are found for [Y/Fe] vs. [La/Fe] and [Zr/Fe] vs. [Ba/Fe], where La
and Ba are measures of the true $r$-contributions. These results 
clearly show that there are major contributions from the third source producing 
Fe with no CPR elements and no $r$-nuclei. Thus, according to the
neutrino-driven wind model for production of the CPR elements, this third
source cannot be massive stars ending as neutron stars but those producing
black holes.

The matter at hand is: What is the nature of this third source? Consideration of 
the yields of VMS ($\sim 140$--$260\,M_\odot$) associated with 
pair-instability SNe (PI-SNe) shows that these sources are characterized by 
strong deficiencies in the elements of odd atomic numbers (e.g., Na, Al, K, Sc, 
V, Mn, Co). Neither the data from earlier studies by \citet{mcw}
nor the more precisely-determined data from recent studies by \citet{cayrel}
on low-metallicity halo stars exhibit these deficiencies. It follows that PI-SNe
cannot be the third source. A plausible candidate is hypernovae (HNe) from
progenitors of $\sim 25$--$50\,M_\odot$. These stars are known to be active 
in the current epoch, although little attention has been paid to them in 
consideration of GCE. They have explosion energies 
far above those of low-mass and normal SNe II and are presumed to be associated 
with gamma-ray bursts. The yields of HNe are generally not well known, but the 
typical Fe yield inferred from observations is $\sim 0.5\,M_\odot$,
much higher than the yield of $\sim 0.07\,M_\odot$ for normal SNe II 
\citep{tominaga}.

\begin{figure}
\begin{center}
\includegraphics[angle=270,scale=0.3]{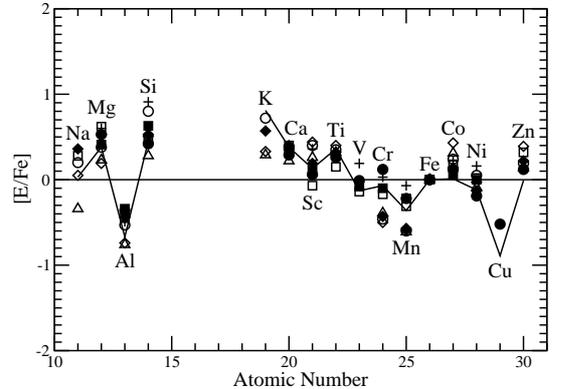}
\caption{Data on the low-$A$ elements from Na through Zn for stars with
``pure'' HN contributions ($f_{{\rm Fe},L}=0$ in Figure~\protect\ref{fig4}b). For
reference, the solid curve represents the abundance pattern measured
for a star with [Fe/H]~$=-2$, which was previously identified with the yield pattern
of the low-$A$ elements for normal SNe II ($L$). There is no apparent difference
between this pattern and the data points for stars with no normal SN II contributions.
\label{fig5}}
\end{center}
\end{figure}

If we assume a Salpeter initial mass function (IMF) for massive stars of 
$\sim 8$--$50\,M_\odot$, the relative rates are $R_{\rm HN}:R_H:R_L
\sim 0.36:0.96:1$ for HNe, low-mass SNe II ($H$), and normal SNe II ($L$), respectively.
Of the Fe contributed by massive stars to the ISM/IGM, the fraction from HNe is
$\sim 0.72$. Thus HNe are the dominant Fe source at early epochs with contributions
far exceeding those of normal SNe II. This seems to explain the earlier conundrum 
\citep{qw02} that the yield patterns for the low-$A$ elements 
from Na through Zn in stars at very low 
metallicities ([Fe/H]~$<-3$) appear to be indistinguishable from what was attributed to 
normal SNe II at higher metallicities (see Figure~\ref{fig5}). This general regularity of
abundance patterns was found by \citet{mcw} and shown more extensively and precisely
by \citet{cayrel}. Figure~\ref{fig5} clearly shows that even for stars deficient in Sr, Y, or Zr,
the abundance patterns for all the low-$A$ elements are approximately constant. 
Thus the diverse normal SN II contributions
cannot govern these abundance patterns. The corresponding
abundances of the low-$A$ elements must then reflect the dominant input to
the ISM/IGM from HNe and not normal SNe II.

\begin{figure}
\begin{center}
\includegraphics[angle=270,scale=0.3]{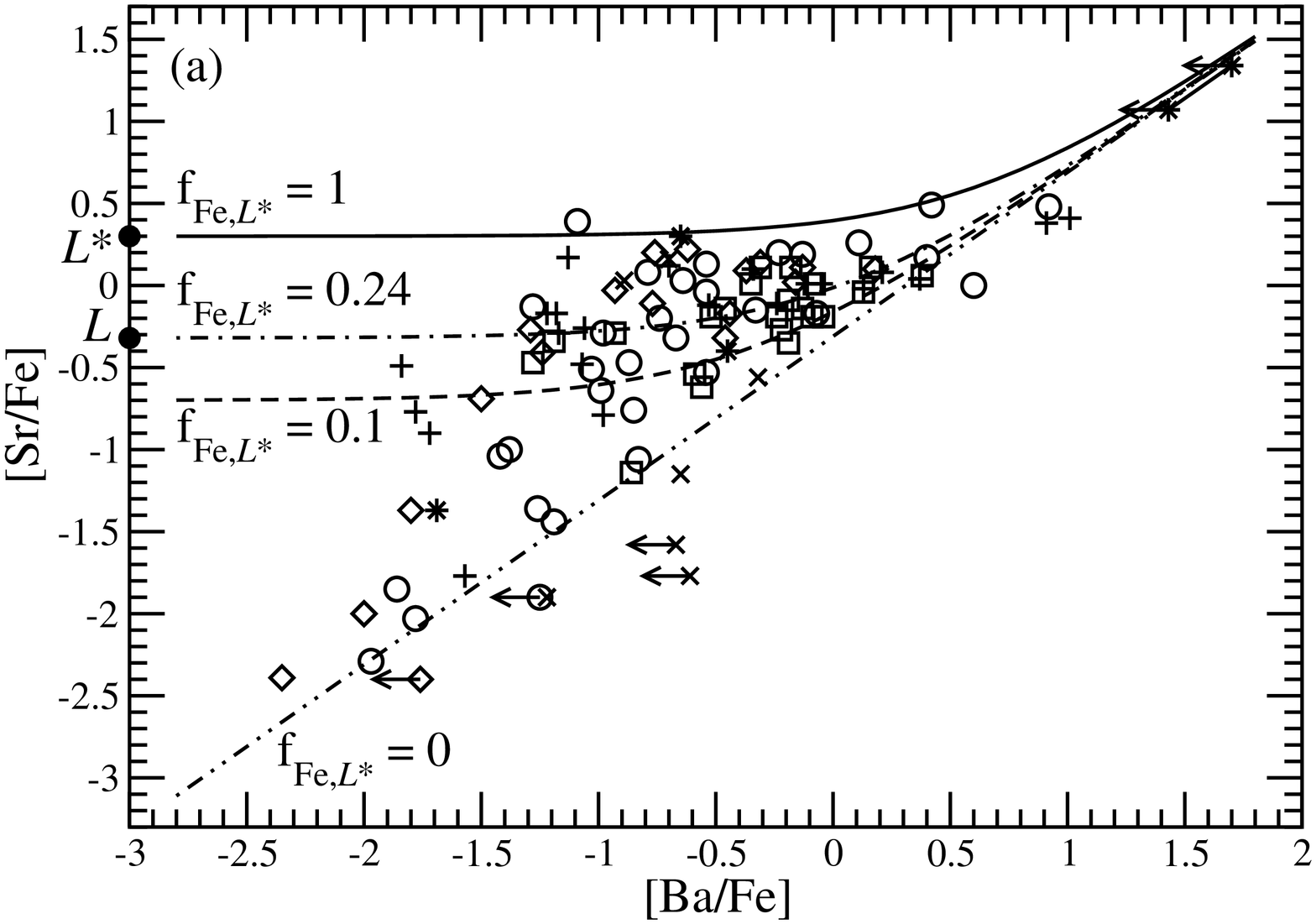}
\includegraphics[angle=270,scale=0.3]{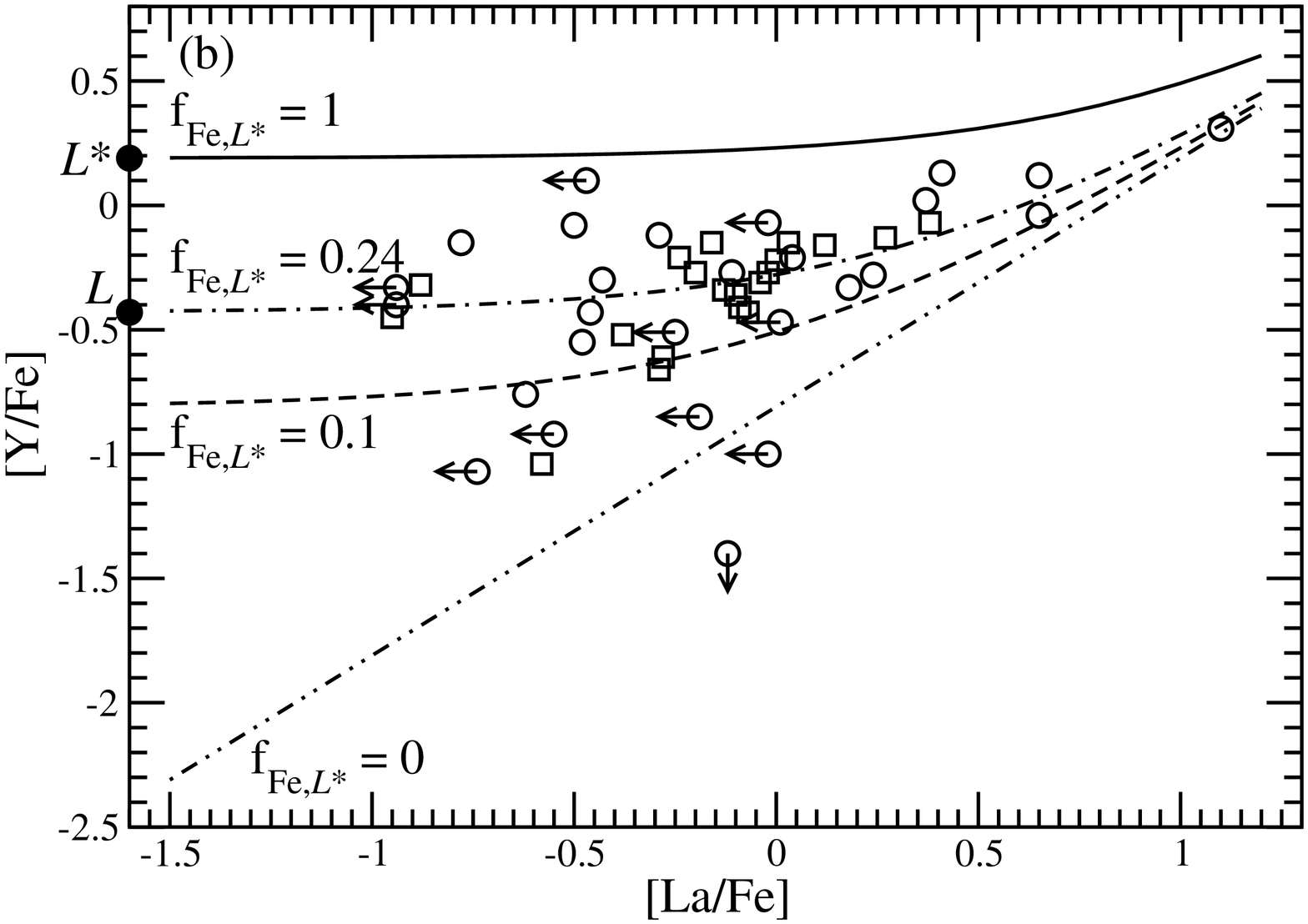}
\caption{(a) Similar to Figure~\protect\ref{fig4}b but with the $L$ source being
a blend of normal SNe II ($L^*$) and HNe (i.e., $L=L^*+{\rm HNe}$). 
(b) Same as (a) but for [Y/Fe] vs. [La/Fe].
The $L^*$ yield ratios (Sr/Fe)$_{L^*}$ and (Y/Fe)$_{L^*}$ are higher
than the corresponding $L$ yield ratios in the two-component model as the
latter ratios include the contributions from HNe producing Fe but no CPR elements.
Note that there are no data points lying above the $f_{{\rm Fe},L^*}=1$ curves
(cf. Figure~\protect\ref{fig4}b). Only one data point lies below the
$f_{{\rm Fe},L^*}=0$ curve for [Y/Fe] vs. [La/Fe].\label{fig6}}
\end{center}
\end{figure}

It is now evident that the inventory of the low-$A$ elements including Fe that we 
had attributed to normal SNe II is in fact the mixture of HN and normal SN II
ejecta where the preponderant contributions are from HNe. The stars that
formed in the juvenile universe appear to have inherited the bulk
of their low-$A$ elements from the ISM where the dominant contributing source
is HNe and not normal SNe II. Thus the 
previously-designated $L$ source is not a pure SN II source but a blend of HNe and 
normal SNe II: $L=L^*+\ {\rm HNe}$, where $L^*$ represents normal SNe II. 
For the estimated yields and relative rates given above, it follows that the 
$\sim 1/3$ of the solar Fe inventory previously assigned to normal SNe II is in 
considerable error. Taking $\sim 2/3$ of the solar Fe inventory to be from
SNe Ia,  we find that of the remaining $\sim 1/3$,  $\sim 24$\% is from HNe with 
only $\sim 9$\% from normal SNe II. Thus with the proper assignment of Fe 
contributions for the $L$ blend with [Sr/Fe]$_L=-0.32$, we obtain 
[Sr/Fe]$_{L^*}=0.3$ for the $L^*$ source. Using the proper yield ratios for the 
$L^*$ source (see Table~3 in \citealt{qw08})  
and equations similar to Equation~(\ref{eq3}), 
we obtain the results shown in Figure~\ref{fig6}. 
It is seen that the data for all the elements appear to be described very well
by the three-component ($H$, $L^*$, and HNe) model. We further note that 
the clump of data above the $f_{{\rm Fe},L} =1$ curve in Figure~\ref{fig4}b 
are now absent in Figure~\ref{fig6}a with diminished Fe contribution from 
normal SNe II. It is also important to note that as amply testified to by 
gamma-ray bursts (e.g., \citealt{grb}), 
HNe are active in the present epoch. Consequently, it is evident that HNe
have been ongoing major contributors to the chemical evolution of the ISM/IGM
during and beyond the early epochs.
The stellar sources for the three-component model are summarized in
Table~\ref{tab2}.

\begin{table}[h]
\begin{center}
\caption{Three-Component Model}\label{tab2}
\begin{tabular}{lccc}
\hline & low-$A$& CPR& heavy \\
&elements&elements&$r$-elements\\
\hline 
low-mass&no&yes&yes\\
SNe II ($H)^a$&&&\\
normal&yes&yes&no\\
SNe II ($L^*)^a$&&&\\
HNe&yes&no&no\\
\hline
\end{tabular}
\medskip\\
$^a$[Sr/Ba]$_H=-0.31$, [Sr/Fe]$_{L^*}=0.3$.\\
\end{center}
\end{table}

\section{General discussion}
From the results reviewed above it appears that the whole chemical evolution 
in the ``juvenile epoch'' of the first $\sim 10^9$~yr after the Big Bang, 
which was dominated
by massive stars, may be explained by 
the concurrent contributions of massive stars associated with low-mass and
normal SNe II and HNe with a standard IMF and that this same relative contribution 
continues into the present epoch. It is possible that the mass range could go up
to $\sim 100\,M_\odot$ (e.g., \citealt{whw}).
The efforts to seek Pop III stars that only occur 
in early epochs and then stop are considered by us to be invalid as were our 
earlier efforts to find a $P$-inventory in the ISM/IGM. It follows that models
for the formation of the ``first'' stars, which has been the focus of intensive studies 
with due consideration of the complex condensation and cooling processes at 
zero to low metallicities (e.g., \citealt{abel02,bromm04}), should consider the stellar 
populations inferred here with HNe ($\sim 25$--$50\,M_\odot$, possibly
up to $\sim 100\,M_\odot$) being the dominant metal source. 
It is possible that more massive stars may have occurred during the very early
stages, but their contributions to the ISM/IGM are quite small at
$-5.5<{\rm [Fe/H]}<-3$. HN explosions are
highly disruptive and certainly can disperse debris through the IGM until halos of 
substantial mass have formed. The apparent 
sudden onset of heavy $r$-elements, which motivated our earlier 
search for a $P$-inventory, is most plausibly related to the formation of 
halos of sufficient mass that remain bound following both SN II and HN explosions 
\citep{qw04}. 
We no longer consider our hypothesis of a $P$-inventory to be valid.
It also follows that the earlier models of GCE that aimed to
provide $\sim 1/3$ of the solar Fe inventory by normal SNe II must now be subject to 
reinvestigation. The observational evidence for ongoing HNe in the current epoch 
cannot be ignored. 

There is further the fact that production of heavy (true) 
$r$-nuclei is strongly decoupled from Fe production. It is also worth noting that
quantitative yields of the CPR elements have not been obtained.  
These issues and the true site of the $r$-process itself are not resolved and
present a further challenge to future stellar models. Insofar as the proposed 
phenomenological three-component model is concerned, more high-resolution
data on the CPR elements (e.g., Sr, Y, and Zr) and heavy $r$-elements (e.g., Ba
and La) are needed to provide a stricter test. It is extremely important to obtain
data for those stars that appear to represent mixtures of $H$
and HN contributions only ($f_{{\rm Fe},L^*}=0$).

With regard to the $r$-process, certainly the possibility that shocked surface layers 
of the core in low-mass SNe II may experience rapid expansion to enable the
production of heavy $r$-elements must be investigated further \citep{ning}.
As there is only limited knowledge on the complex evolution of the progenitors
of $\sim 8$--$11\,M_\odot$ for low-mass SNe II, more extensive and intensive 
studies of these stars are required. The problem of explaining the meteoritic data
on $^{129}$I and $^{182}$Hf, which launched the investigation into multiple
$r$-processes, is still not resolved. As emphasized by K.-L. Kratz
(e.g., \citealt{ott}), from the
point of view of nuclear systematics, there is substantial difficulty in
understanding a break in the production between these two nuclei (but see
\citealt{qvw}). These and other issues discussed above provide ample stimulus
for future studies of the $r$-process in particular and GCE in general.
In the phenomenological approach used by us, it is assumed that stars 
themselves have correctly calculated the yields of stellar nucleosynthesis. 
Theory is the important guide in illuminating the path to understanding.

\section*{Acknowledgments} %If needed
This paper is affectionately dedicated to our friend Roberto Gallino, who has led in
our exploration and understanding of
what happens in stellar nucleosynthesis in low-mass
stars. We have wandered together with him into unknown dimensions of multiple 
``$r$-processes.'' Sometimes, we are led into real surprises and some new insights.
This work was supported in part by US DOE grant DE-FG02-87ER40328 (Y.Z.Q.).
G.J.W. acknowledges NASA's Cosmochemistry Program for research support 
provided through J. Nuth at the Goddard Space Flight Center. He also
appreciates the generosity of the Epsilon Foundation.

%\end{multicols}

\end{document}